\documentclass{article}
\usepackage{spconf,amsmath,graphicx}

\usepackage[final]{microtype}
\usepackage[square, sort, comma, numbers, compress]{natbib}

\usepackage{enumitem}
\usepackage{booktabs}
\usepackage{amsfonts}
\usepackage{bm}
\usepackage{array}
\newcolumntype{C}[1]{>{\centering\arraybackslash}p{#1}}
\setlist{nosep, leftmargin=14pt}
\usepackage[colorlinks=true]{hyperref}

\usepackage{mwe} 

\def\Tone{{T$_1$-w}}

\title{Optimal operating MR contrast for brain ventricle parcellation}
\address{%
$^{1}$ Department of Electrical and Computer Engineering, Johns Hopkins University, USA\\
$^{2}$ Laboratory of Behavioral Neuroscience, National Institute on Aging,\\ National Institutes of Health, USA\\
$^{3}$ Department of Biomedical Engineering, Johns~Hopkins~School~of~Medicine, USA\\
$^{4}$ Department of Neurosurgery, Johns~Hopkins~School~of~Medicine, USA}

\begin{document}
%
\maketitle
Development of MR harmonization has enabled different contrast MRIs to be synthesized while preserving the underlying anatomy.
In this paper, we use image harmonization to explore the impact of different \Tone~MR contrasts on a state-of-the-art ventricle parcellation algorithm VParNet.
We identify an optimal operating contrast~(OOC) for ventricle parcellation; by showing that the performance of a pretrained VParNet can be boosted by adjusting contrast to the OOC.

\vspace*{1em}
\noindent{\large{\textbf{Introduction and Methods}}}

\label{sec:intro}
Harmonization can be used to evaluate the impact of MR image contrast of \Tone~images on brain segmentation~\cite{hays12032evaluatingS}. 
In this paper, we propose a framework to improve the applicability and accuracy of the current state-of-the-art ventricle parcellation method, \underline{VParNet~\cite{shao2019nicS}, without retraining the model}. 

\begin{figure}[!tb]
    \centering
    \includegraphics[width=0.75\columnwidth]{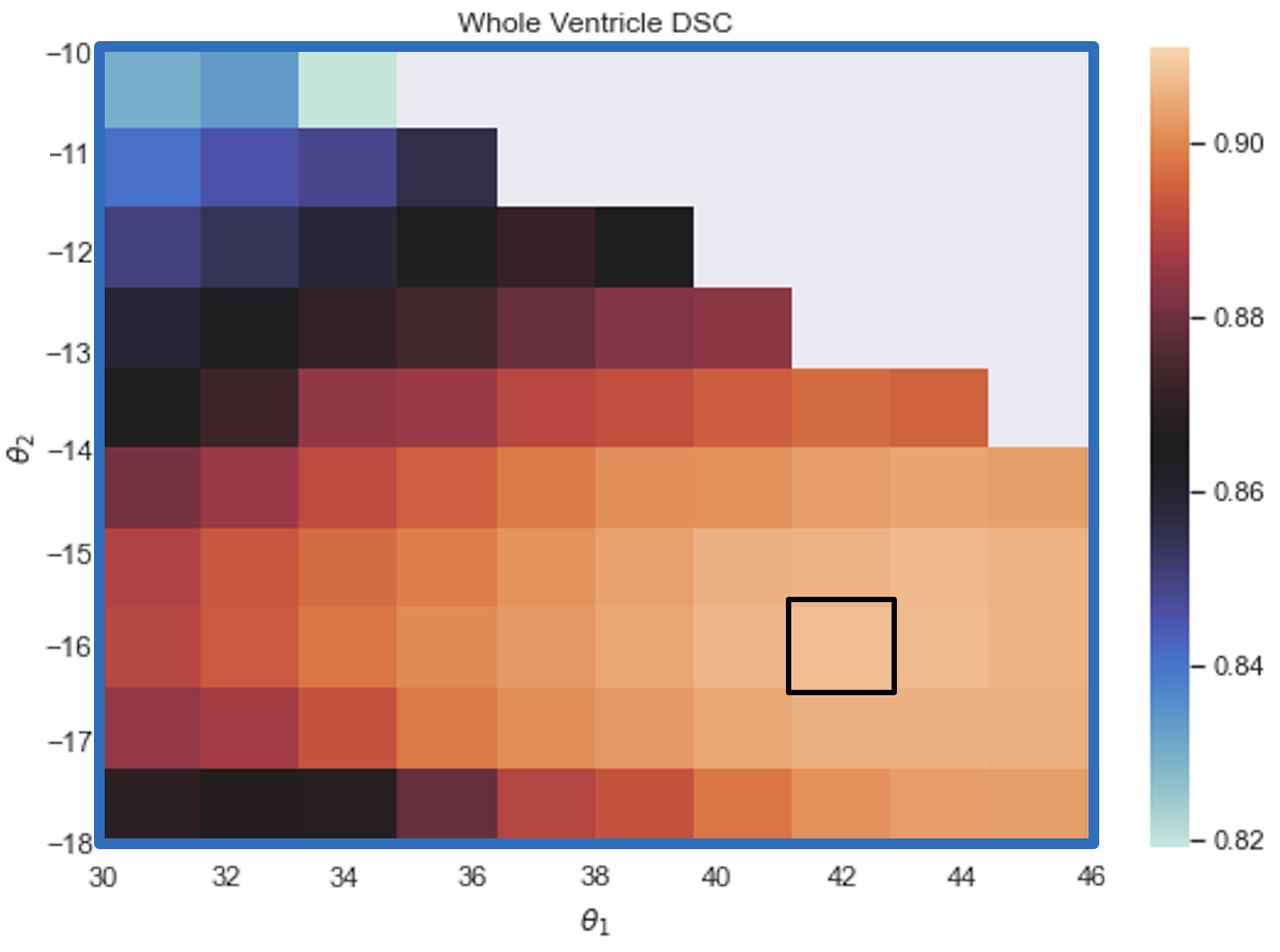}
    \caption{OOC of VParNet identified by a grid search of the CALAMITI contrast space. Dice based heatmap indicates ventricle parcellation performance of VParNet on eight sample subjects.}
    \label{fig:heatmap}
\end{figure}

Using CALAMITI ~\cite{zuo2021unsupervisedS}, we generate synthetic MR images with contrast $\bm{\theta}$ but the same underlying anatomy. 
To find the OOC for VParNet, we considered 100 different contrasts within the blue rectangle shown in Fig.~\ref{fig:heatmap}. 
We rejected 21 locations as their contrast does not resemble~\Tone{}eighting. 
The synthetic images and corresponding ventricle delineations were used to evaluate VParNet performance across the different contrasts. 
We identified an OOC of VParNet by selecting the target contrast with the highest mean Dice with respect to the ventricle delineations shown in Fig.~\ref{fig:heatmap}.

\vspace*{1em}
\noindent{\large{\textbf{Experiments and Results}}}

\label{sec:experiments}
We adjusted the contrast of an additional $35$ subjects that were not included in VParNet training to the OOC. 
A paired Wilcoxon signed rank test
was conducted on the Dice between the ventricle parcellation results on the original images and the contrast adjusted images.
For the whole ventricle label, VParNet has Dice of 0.886 $\pm$ 0.04 on the original images, while VParNet has Dice of 0.895$\pm$0.03 on the images adjusted to the OOC.
There was a significant ($p<0.01$) improvement between VParNet performance on the original and OOC adjusted images for the 4th, left and right lateral, and whole ventricle.


\vspace*{1em}
\noindent{\large{\textbf{Discussion and Conclusion}}}
\label{sec:dis}

In this work, we demonstrated improved VParNet performance by adjusting input image contrast using harmonization~(and no retraining).
We successfully identified an OOC of a pretrained VParNet with a grid search in contrast space.
After contrast adjustment, we demonstrate improvement in VParNet performance on data that is from the same cohort as VParNet training data. 
On the contrary, we identify areas of contrast space where VParNet may perform poorly. 

\bibliographystyle{IEEEbib}
\bibliography{strings,refs}

\end{document}